\documentclass[aps,10pt,prd,twocolumn,floats,letterpaper,showpacs,nofootinbib,notitlepage,superscriptaddress]{revtex4-1}

\usepackage{graphicx}
\usepackage{mathrsfs}
\usepackage[intlimits,centertags]{amsmath}
\usepackage{amssymb,amsfonts}
\usepackage[pdftex]{hyperref}
\usepackage[x11names]{xcolor}
\usepackage{mathpazo}
\usepackage{enumerate}
\usepackage{xfrac}
\usepackage[normalem]{ulem}

\hypersetup{pdftitle={Unitarity Bounds of Astrophysical Neutrinos},
pdfsubject={Unitarity Bounds of Astrophysical Neutrinos},
pdfauthor={Markus Ahlers, Mauricio Bustamante, and Siqiao Mu},
pdfstartview={FitH},
colorlinks=true,
bookmarksopen=false,
bookmarksnumbered=false,
bookmarksopenlevel=0,
linkcolor=Blue1!60!black,
citecolor=Green1!50!black,
urlcolor=Blue1!70!black
}

\begin{document}

\title{Unitarity Bounds of Astrophysical Neutrinos}
\author{Markus Ahlers}
\email{markus.ahlers@nbi.ku.dk}
\thanks{ORCID: \href{http://orcid.org/0000-0003-0709-5631}{0000-0003-0709-5631}}
\affiliation{Niels Bohr International Academy \& Discovery Centre, Niels Bohr Institute,\\University of Copenhagen, Blegdamsvej 17, DK-2100 Copenhagen, Denmark}
\author{Mauricio Bustamante}
\email{mbustamante@nbi.ku.dk}
\thanks{ORCID: \href{http://orcid.org/0000-0001-6923-0865}{0000-0001-6923-0865}}
\affiliation{Niels Bohr International Academy \& Discovery Centre, Niels Bohr Institute,\\University of Copenhagen, Blegdamsvej 17, DK-2100 Copenhagen, Denmark}
\affiliation{DARK, Niels Bohr Institute, University of Copenhagen, Blegdamsvej 17, DK-2100 Copenhagen, Denmark}
\author{Siqiao Mu}
\email{smu@caltech.edu}
\affiliation{California Institute of Technology, 1200 E California Blvd, Pasadena, CA 91125, USA}
\begin{abstract}
The flavor composition of astrophysical neutrinos observed at neutrino telescopes is related to the initial composition at their sources via oscillation-averaged flavor transitions. If the time evolution of the neutrino flavor states is unitary, the probability of neutrinos changing flavor is solely determined by the unitary mixing matrix that relates the neutrino flavor and propagation eigenstates. In this paper we derive general bounds on the flavor composition of TeV--PeV astrophysical neutrinos based on unitarity constraints. These bounds are useful for studying the flavor composition of high-energy neutrinos, where energy-dependent nonstandard flavor mixing can dominate over the standard mixing observed in accelerator, reactor, and atmospheric neutrino oscillations. 
\end{abstract}

\pacs{14.60.Pq, 14.60.St, 95.55.Vj}

\maketitle

\section{Introduction}

The high-energy astrophysical neutrinos discovered by IceCube~\cite{Aartsen:2013bka,Aartsen:2013jdh,Aartsen:2014gkd,Aartsen:2015rwa,Aartsen:2016xlq,IceCube:2018dnn,IceCube:2018cha} are key to revealing the unknown origin of high-energy cosmic rays and the physical conditions in their sources~\cite{Ahlers:2018fkn}. They can escape dense environments, that are otherwise opaque to photons, and travel cosmic distances without being affected by background radiation or magnetic fields. They also provide a unique opportunity to study fundamental neutrino properties in an entirely new regime: their energy and baseline far exceed those involved in reactor, accelerator, and atmospheric neutrino experiments. Effects of nonstandard neutrino physics --- even if they are intrinsically tiny --- can imprint themselves onto the features of astrophysical neutrinos, including their energy spectrum, arrival directions, and flavor composition, {\it i.e.,} the proportion of neutrinos of each flavor.

\begin{figure}[b]\centering
\includegraphics[width=0.95\linewidth,viewport=40 50 400 375,clip=true]{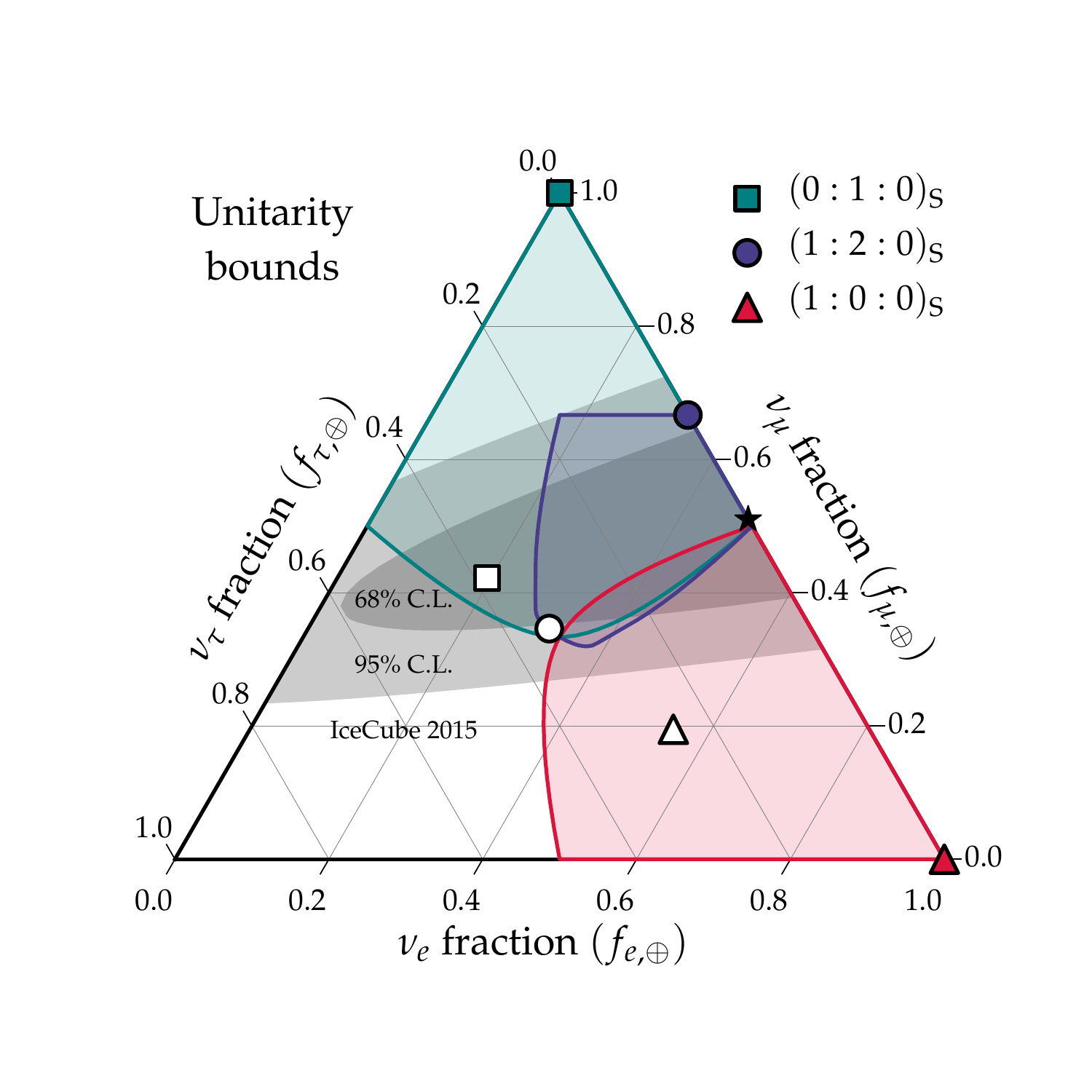}
\caption[]{Unitarity bounds of astrophysical neutrino flavors for three source compositions indicated by filled symbols. The corresponding open symbols indicate the expected composition at Earth under standard oscillations using the best-fit mixing parameters for normal mass ordering~\cite{Esteban:2016qun}. We include the best-fit flavor composition from IceCube~\cite{Aartsen:2015knd} as a black star and the $68\%$ and $95\%$ confidence levels as grey-shaded areas.}\label{fig1}
\end{figure}

At the sources, high-energy neutrinos ($\gg {\rm GeV}$) are produced by cosmic-ray interactions with gas and radiation. These neutrinos are flavor eigenstates from the weak decay of secondary particles. The initial composition of neutrino flavor states is determined by details of the production process. After emission, oscillations modify the composition en route to Earth~\cite{Beacom:2003nh, Kashti:2005qa, Xing:2006uk, Lipari:2007su, Pakvasa:2007dc, Esmaili:2009dz, Lai:2009ke, Choubey:2009jq}. Assuming standard oscillations, we can predict the observable flavor composition from a given source composition. However, nonstandard neutrino oscillations can alter the composition drastically~\cite{Bustamante:2010bf, Xu:2014via, Fu:2014isa, Bustamante:2015waa, Gonzalez-Garcia:2016gpq, Nunokawa:2016pop, Rasmussen:2017ert}. Nonstandard effects can originate, {\it e.g.}, from neutrino interactions with background matter~\cite{Bustamante:2018mzu}, dark matter~\cite{deSalas:2016svi,Capozzi:2018bps} or dark energy~\cite{Klop:2017dim,Capozzi:2018bps} or from Standard Model extensions that violate the weak equivalence principle, Lorentz invariance, or CPT symmetry~\cite{DeSabbata:1981ek, Gasperini:1989rt, Glashow:1997gx, Barenboim:2003jm, Bustamante:2010nq, Esmaili:2014ota, Arguelles:2015dca, Lai:2017bbl}. A key property of these models is that the flavor transitions between sources and Earth are entirely determined by a new unitary mixing matrix that connects neutrino flavor and propagation eigenstates~\cite{Akhmedov:2017mcc}.

We will discuss the regions in flavor space that can be expected from this class of models. The unitarity of the new mixing matrix allows us to compute the boundary of the region that encloses all possible flavor compositions at the Earth, in spite of not knowing the values of the matrix elements. Previous work~\cite{Xu:2014via} derived a set of unitarity bounds for specific choices of flavor composition at the sources. We extend this work by providing a refined and explicit formalism to derive unitarity bounds that are easily applicable to arbitrary source compositions.

Figure \ref{fig1} shows our results for physically motivated choices of source flavor composition. The ternary plot shows the source and Earth flavor fractions, {\it i.e.}, the relative contribution of neutrino flavors to the total neutrino flux. Assuming that the accessible flavor space is convex, {\it i.e.}, that every intermediate flavor fraction between any two accessible fractions is also accessible by a suitable unitary matrix, our unitarity bounds are maximally constraining and completely characterize the accessible flavor space. 

The paper is organized as follows. In Sec.~\ref{sec1} we discuss the astrophysical processes of neutrino production and the corresponding flavor composition at the source. We discuss the resulting flavor composition at Earth after flavor oscillation with nonstandard neutrino mixing. In Sec.~\ref{sec2} we derive general flavor boundaries for the flavor composition at Earth based on the unitary of the nonstandard mixing matrix. We conclude in Sec.~\ref{sec3}. Throughout this paper we will work in natural units with $\hbar=c=1$.

\section{Astrophysical Neutrino Flavors}\label{sec1}

High-energy astrophysical neutrinos are products of cosmic-ray collisions with gas and radiation. The flux of neutrinos at production can be described as a mixed state of neutrinos $\nu_\alpha$ and antineutrinos $\overline\nu_\alpha$ where the index $\alpha = e, \mu, \tau$ refers to the neutrino flavor eigenstate produced in weak interactions. The relative number of initial neutrino states $(N_e$\,:\,$N_\mu$\,:\,$N_\tau)_{\rm S}$ (summed over neutrinos and antineutrinos) is determined by the physical conditions in the source. In the simplest case, pions (or kaons) produced in cosmic-ray interactions decay via $\pi^+\to\mu^++\nu_\mu$ followed by $\mu^+\to e^++\nu_e+\overline\nu_\mu$ (and the charge-conjugated processes). This pion decay chain results in a source composition of $($1\,:\,2\,:\,0$)_{\rm S}$. However, in the presence of strong magnetic fields it is possible that muons lose energy before they decay and do not contribute to the high-energy neutrino emission~\cite{Kashti:2005qa}. In this muon-damped scenario the composition is expected to be closer to $($0\,:\,1\,:\,0$)_{\rm S}$. On the other hand, neutrino production by beta-decay of free neutrons or short-lived isotopes produced in spallation or photo-disintegration of cosmic rays leads to $($1\,:\,0\,:\,0$)_{\rm S}$.

After production, astrophysical neutrinos travel over cosmic distances before their arrival at Earth. The observable flavor composition is significantly altered by neutrino oscillations, which are due to neutrino flavor states being superpositions of propagation states $\nu_\mathfrak{a}$,
\begin{equation}\label{eq:Uneutrino}
  |\nu_\alpha\rangle =
  \sum_\mathfrak{a} U_{\alpha \mathfrak{a}}^* |\nu_\mathfrak{a}\rangle\,.
\end{equation}
These propagation states are defined as eigenvectors of the Hamiltonian, including kinetic terms and effective potentials~\cite{Akhmedov:2017mcc}. In general, the $3\times3$ unitary mixing matrix $\bf U$ has nine degrees of freedom. However, neutrino oscillation phenomena only depend on four independent parameters, which can be parametrized by three rotation angles and one phase. Unitarity ensures that the total number of neutrinos of all flavors is conserved. Neutrino flavor oscillations of pure or mixed states can be described in terms of the evolution of the density matrix $\rho$, following the Liouville equation $\dot\rho= -{\rm i}[H,\rho]$ with Hamiltonian $H$.

In the case of standard neutrino oscillations and neutrino propagation in vacuum, the propagation eigenstates are identical to the neutrino mass eigenstates $\nu_i$ ($i=1,2,3$). The mixing matrix between flavor and mass eigenstates is the so-called {\it Pontecorvo-Maki-Nagakawa-Sakata} (PMNS) matrix~\cite{Pontecorvo:1957qd,Maki:1962mu,Pontecorvo:1967fh}. In the relativistic limit, standard oscillations in vacuum can be introduced via the Hamiltonian
\begin{equation}\label{eq:H0}
  H_{0} \simeq  \sum_i\frac{m_i^2}{2E_\nu}\big(|\nu_i\rangle\langle\nu_i|+|\overline{\nu}_i\rangle\langle\overline{\nu}_i|\big)\,,
\end{equation}
where $E_\nu$ is the neutrino energy and the sum runs over projectors onto neutrino and antineutrino mass eigenstates.

The solution of the Liouville equation with $H = H_0$ describes the oscillation of neutrino flavors due to the nontrivial mixing and mass splitting, $\Delta m^2 \equiv m_i^2 - m_j^2 \neq 0$, for $i \neq j$. The oscillation phases are given by $\Delta m^2\ell/4E_\nu$ where $\ell$ is the distance to the neutrino source. In the case of astrophysical neutrinos these oscillation phases are much larger than unity. Considering the wide energy distribution of neutrinos at their sources and the limited energy resolution of neutrino detectors, flavor transitions from $|\nu_\alpha\rangle$ to $ |\nu_\beta\rangle$ (or from $|\overline\nu_\alpha\rangle$ to $ |\overline\nu_\beta\rangle$) can only be observed by their oscillation-averaged transition probability given by
\begin{equation}\label{eq:Paverage}
{P}_{\alpha\beta} = \sum_\mathfrak{a}|U_{\alpha \mathfrak{a}}|^2\, |U_{\beta \mathfrak{a}}|^2\,.
\end{equation}

In the following, we will discuss nonstandard neutrino oscillations that can be described by additional effective Hamiltonian terms $\widetilde{H}$ in the Liouville equation, so that the total Hamiltonian is $H=H_0+\widetilde{H}$. These effective terms can be generated in various ways, including nonstandard interactions with matter and Standard Model extensions that violate the weak equivalence principle, Lorentz invariance, or CPT symmetry. Concretely, we will study the effect of additional terms in the Hamiltonian that can be parametrized in the form~\cite{GonzalezGarcia:2005xw}
\begin{equation}\label{eq:Heff}
\widetilde{H} =  \frac{E_\nu^n}{\Lambda^n} \sum_{\mathfrak{a}} \big(\epsilon_\mathfrak{a}|\nu_{\mathfrak{a}}\rangle\langle\nu_{\mathfrak{a}}| + \overline{\epsilon}_\mathfrak{a}|\overline{\nu}_{\mathfrak{a}}\rangle\langle\overline{\nu}_{\mathfrak{a}}|\big) \,,
\end{equation}
with $n$ an integer and $\Lambda$ the energy scale of the nonstandard effects. The eigenvalues of this additional Hamiltonian, $\epsilon_\mathfrak{a}$ and $\overline\epsilon_\mathfrak{a}$, are required to be nondegenerate, $\Delta\epsilon \equiv \epsilon_\mathfrak{a} - \epsilon_\mathfrak{b} \neq 0$, for $\mathfrak{a}\neq\mathfrak{b}$, in order to induce neutrino oscillations. For simplicity, we will consider CP-even Hamiltonians with $\epsilon_\mathfrak{a}=\overline\epsilon_\mathfrak{a}$ that affect neutrinos and antineutrinos equally.

Neutrino oscillations have been studied extensively with reactor, solar, and atmospheric neutrino experiments. Global data confirms the three-flavor oscillation phenomenology parametrized by the PMNS matrix. This allows to derive bounds on the effective Hamiltonian, Eq.~(\ref{eq:Heff})~\cite{Kostelecky:2011gq,Abbasi:2009nfa,Abe:2014wla,Aartsen:2017ibm}, or more general extensions allowing for, {\it e.g.}, anisotropic contributions along an ordered background field. A general classification of these effective Hamiltonians can be found in Ref.~\cite{Kostelecky:2011gq} and experimental limits have been summarized in Ref.~\cite{Kostelecky:2008ts}. 

The sensitivity reach of oscillation experiments to the coefficients of the effective Hamiltonian can be estimated as
\begin{equation}\label{eq:HeffBound}
\frac{\Delta\epsilon}{\Lambda^n}\ll \frac{10^{-3}\,{\rm eV}^2}{(1\,{\rm TeV})^{n+1}}\,,
\end{equation}
where $\Delta \epsilon$ is the largest splitting of the eigenvalues $\epsilon_\mathfrak{a}$ that we compare against oscillations induced by the atmospheric mass splitting and the energy scale of atmospheric neutrinos, about 1~TeV. At higher energies, the standard Hamiltonian Eq.~(\ref{eq:H0}) becomes smaller due to its $1/E_\nu$ dependence, while the relative size of nonstandard effects with $n \geq 0$ grows.  Thus, the higher the energy, the smaller the nonstandard effects that can be tested. In high-energy astrophysical neutrinos, with TeV--PeV energies, even small contributions bounded by Eq.~(\ref{eq:HeffBound}) may dominate oscillations. For instance, assuming $n=0$, the nonstandard contribution Eq.~(\ref{eq:Heff}) can dominate over the standard Hamiltonian Eq.~(\ref{eq:H0}) by two orders of magnitude. If this is the case, the oscillation-averaged flavor-transition matrix will take on a form analogous to Eq.~(\ref{eq:Paverage}), but with a new unitary mixing matrix describing the mixing between flavor states $\nu_\alpha$ and the nonstandard propagation states $\nu_\mathfrak{a}$. The unitary mixing matrix is not constrained by low-energy neutrino data and can, in principle, have elements with values very different from the PMNS mixing matrix.

Various authors
have studied the effects of nonstandard Hamiltonians on the astrophysical neutrino flavor composition and its compatibility with IceCube observations; see {\it e.g.}~\cite{Rasmussen:2017ert}. Independently of the underlying neutrino physics, the flavor composition at Earth is limited by the unitary mixing between flavor eigenstates and the eigenstates of the Hamiltonian. In the following, we will derive unitarity bounds on the observable flavor composition of astrophysical neutrinos. 

\section{Flavor Boundaries}\label{sec2}

\begin{figure}[t]\centering
\includegraphics[width=0.49\linewidth]{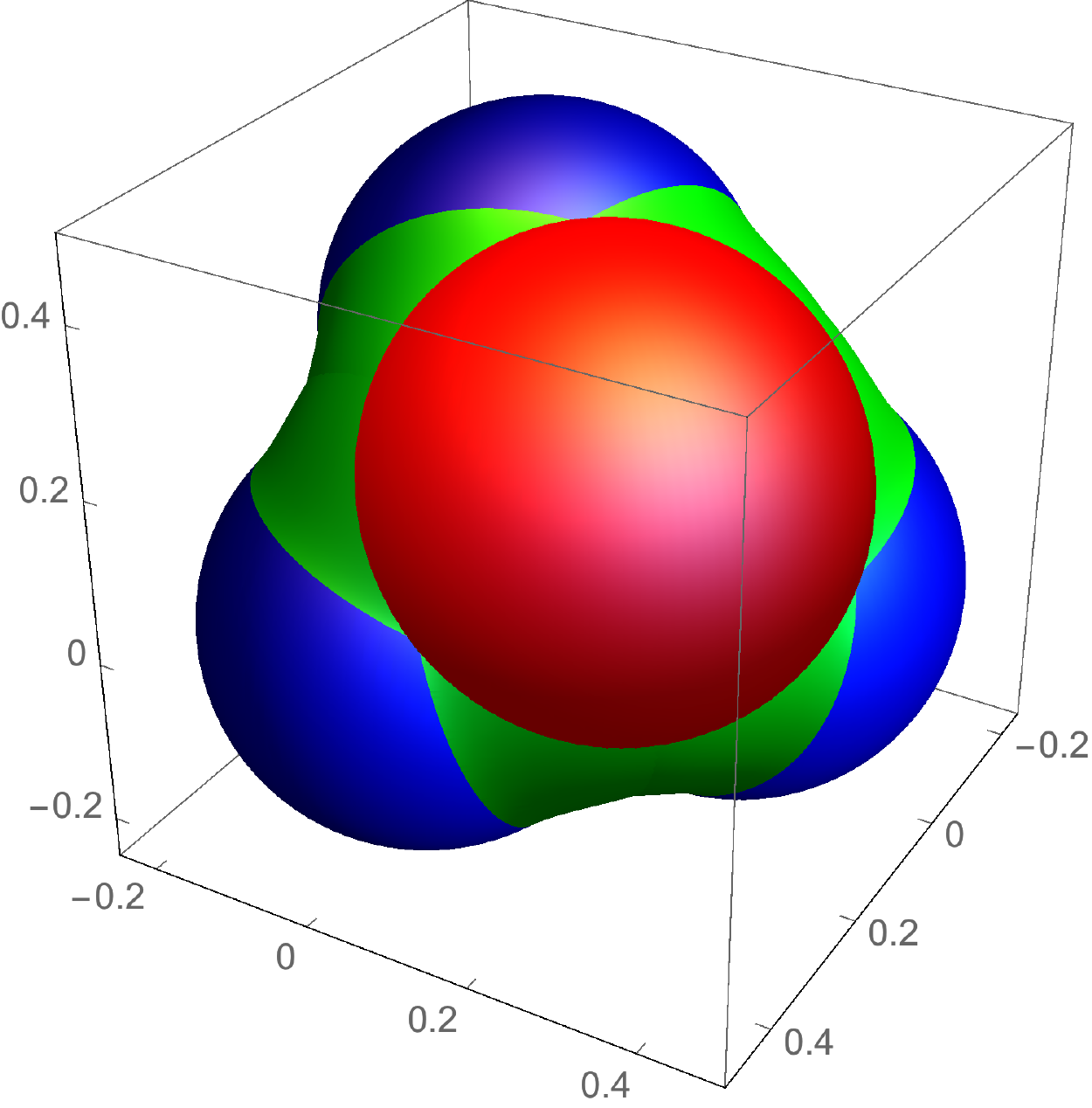}\hfill\includegraphics[width=0.49\linewidth]{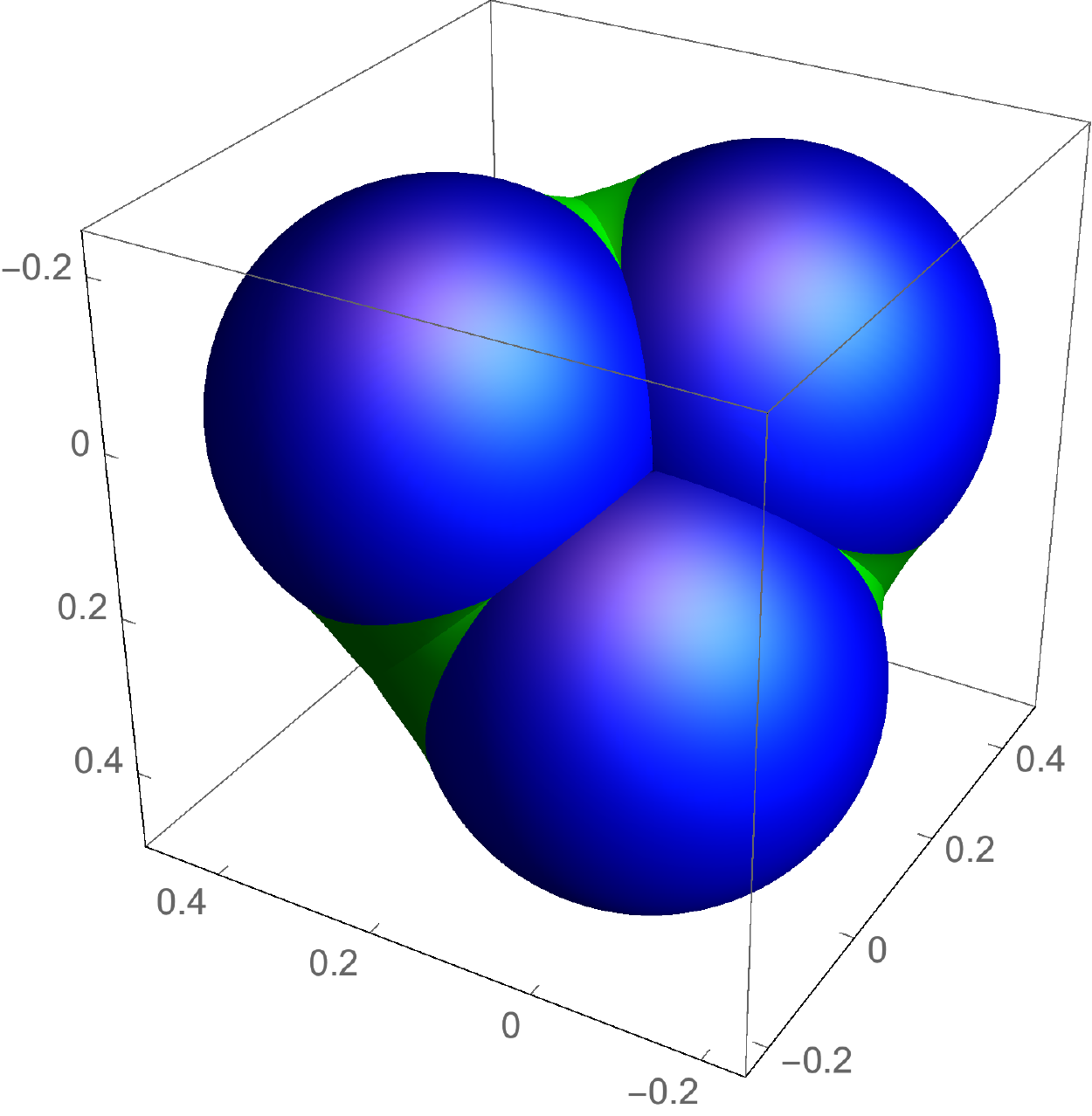}\\
\caption[]{Boundary function $B$ (viewed from opposite directions) parametrized as the surface $B(x,y,z){\bf n}$ with unit vector ${\bf n} = (x,y,z)$. The colors indicate the directions ${\bf n}$ where the boundary is given by the branches $\mathcal{S}_1$ (red), $\mathcal{S}_2$ (blue), or $\mathcal{S}_3$ (green).}\label{fig2}
\end{figure}

The oscillation-averaged flavor transition matrix defined by Eq.~(\ref{eq:Paverage}) can be parametrized by its three off-diagonal entries ${P}_{\mu\tau}={P}_{\tau\mu}$, ${P}_{e\tau} = {P}_{\tau e}$, and ${P}_{e\mu} = {P}_{\mu e}$. The unitarity of the mixing matrix imposes a limit on linear combinations of these transition elements,
\begin{equation}\label{eq:bound}
x{P}_{\mu\tau} + y {P}_{e\tau} + z{P}_{e\mu} \leq B(x,y,z)\,,
\end{equation}
where $x$, $y$, and $z$ are arbitrary parameters. The boundary function $B$ is given by (see Appendix~\ref{appI})
\begin{equation}\label{eq:B}
B(x,y,z) = \max\left( \lbrace0\rbrace\cup\mathcal{S}_1\cup\mathcal{S}_2\cup\mathcal{S}_3\right)\,,
\end{equation}
where the individual subsets $\mathcal{S}_i$ correspond to different branches and are defined as
\begin{gather}
\mathcal{S}_1(x,y,z) = \Big\lbrace\frac{x+y+z}{3}\Big\rbrace\,,
\\
\mathcal{S}_2(x,y,z) = \Big\lbrace\frac{x}{2},\frac{y}{2},\frac{z}{2}\Big\rbrace\,,
\\
\mathcal{S}_3(x,y,z) = \mathcal{S}'(x,y,z)\cup\mathcal{S}'(y,z,x)\cup\mathcal{S}'(z,x,y)\,,
\end{gather}
with
\begin{equation}
\mathcal{S}'(x,y,z) = \Big\lbrace \frac{(3x+y+z)^2-4yz}{24x}\Big|x^2\geq
\frac{(y-z)^2}{9}\Big\rbrace\,.
\end{equation}
Figure~\ref{fig2} shows a graphical representation of $B(x,y,z)$ in terms of the surface $B(x,y,z){\bf n}$ along a unit vector ${\bf n} = (x,y,z)$. The red, blue, and green-colored regions indicate where the different branches $\mathcal{S}_1$, $\mathcal{S}_2$, and $\mathcal{S}_3$, respectively, determine the maximum in Eq.~(\ref{eq:B}).

\begin{figure*}[t]\centering
\includegraphics[width=0.33\linewidth,viewport=40 50 400 375,clip=true]{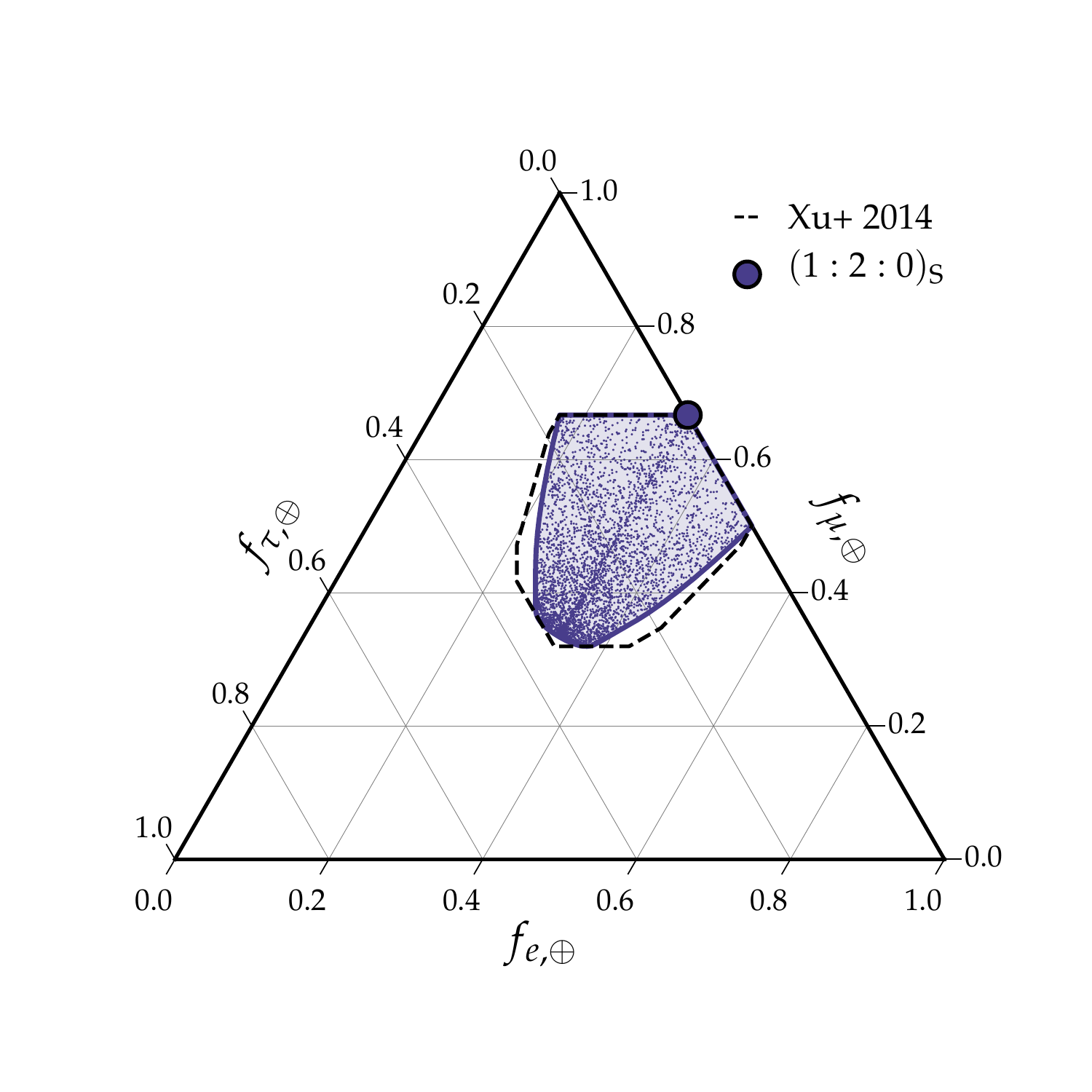}\hfill\includegraphics[width=0.33\linewidth,viewport=40 50 400 375,clip=true]{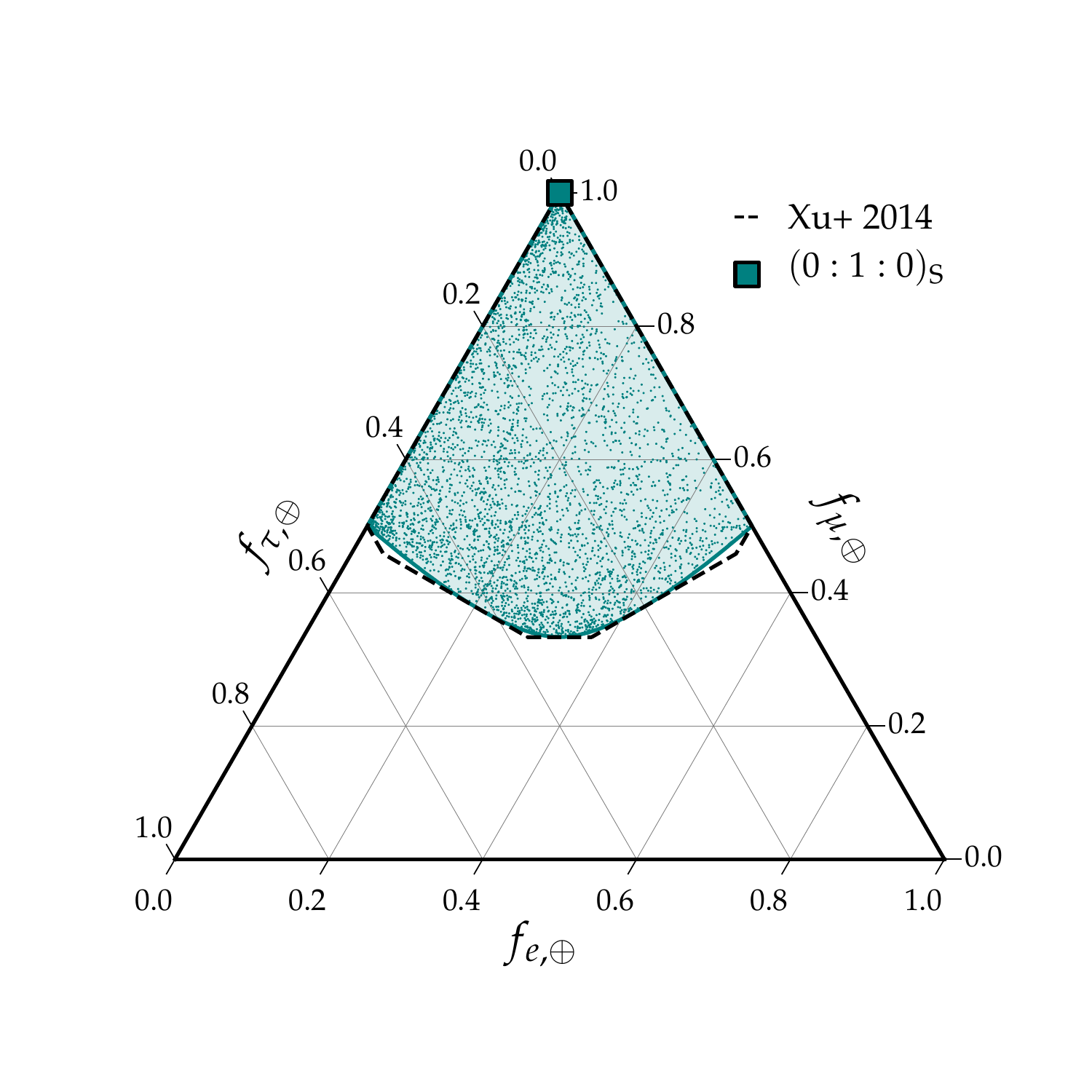}\hfill\includegraphics[width=0.33\linewidth,viewport=40 50 400 375,clip=true]{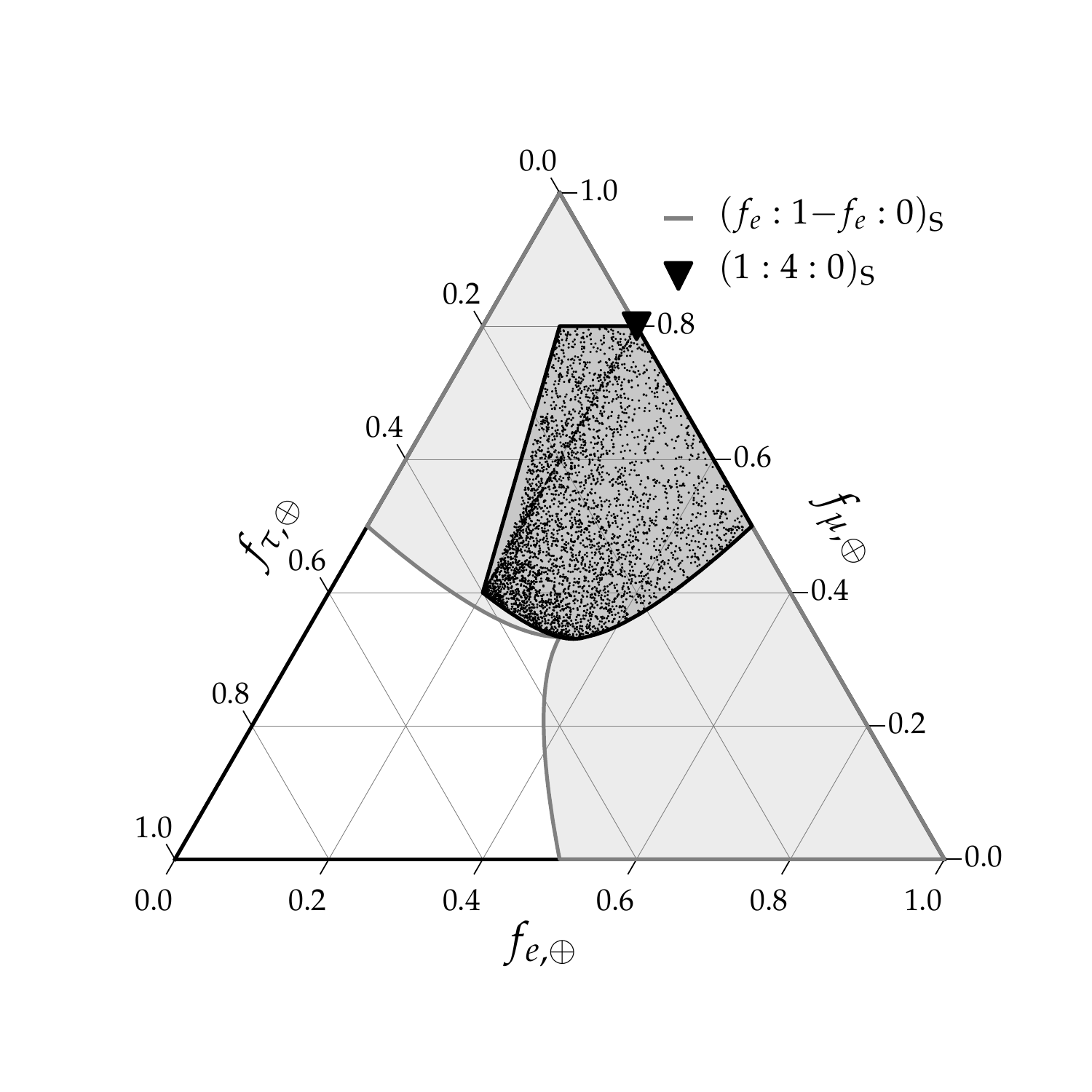}
\caption[]{Comparison of unitarity bounds, Eq.~(\ref{eq:boundary}), to random realizations of the mixing matrix. {\bf Left:} Unitarity bound for a source composition $($1$\,:\,$2$\,:\,$0$)_{\rm S}$, as shown in Fig.~\ref{fig2}, in comparison to 4,000 random samples. We also show the bound derived in Ref.~\cite{Xu:2014via}. {\bf Center:} Same as in the left panel, but for $($0$\,:\,$1$\,:\,$0$)_{\rm S}$. Note that this is related to $($1$\,:\,$0$\,:\,$0$)_{\rm S}$ after index permutations, as described in Appendix~\ref{appI}. {\bf Right:} The general boundary condition for $(f_e$\,:\,$1{\rm -}f_e$\,:\,$0)_{\rm S}$, {\it i.e.}, no $\nu_\tau$ production. We also show the unitary bound for the special case $(1$\,:\,$4$\,:\,$0)_{\rm S}$. For this particular source composition the accessible flavor space appears to be concave and the boundary derived by Eq.~(\ref{eq:boundary}) is not maximally constraining.}\label{fig3}
\end{figure*}

It is possible to use the family of bounds in Eq.~(\ref{eq:G}) of Appendix~\ref{appI} to derive boundaries that enclose the accessible region of observable flavor compositions. We define the flavor ratio as $f_{\alpha} \equiv N_\alpha/\sum_\beta N_\beta$. For a fixed source flavor ratio $f_{\alpha,{\rm S}}$, any observable ratio $f_{\alpha,\oplus}$ has to obey the relation
\begin{equation}
f_{\alpha,\oplus} = \sum_\beta{P}_{\alpha\beta}f_{\beta,{\rm S}}\,.
\end{equation}
For trivial mixing, ${\bf U}={\bf I}$, the oscillation-averaged transition probability is also trivial and $f_{\alpha,\oplus}=f_{\alpha,{\rm S}}$. Therefore, the original flavor composition is always part of the accessible flavor space. Since there is a continuous parametrization of the transition matrix ${\bf P}$ in terms of unitary mixing angles and phases, the area in the accessible flavor space has to be connected (although not necessarily simply connected). Therefore, we will look for the boundary of the flavor shift defined as 
\begin{equation}
\Delta f_{\alpha} \equiv f_{\alpha,\oplus} - f_{\alpha,{\rm S}}\,.
\end{equation}
This shift can be expressed as a linear combination of transition probabilities ${P}_{\mu\tau}$, ${P}_{e\tau}$, and ${P}_{e\mu}$ and is therefore bounded by the family of bounds in Eq.~(\ref{eq:bound}). Due to unitarity, we have $\sum_\alpha \Delta f_{\alpha} =0$ and we can therefore parametrize the total flavor shift by, say, $\Delta f_e$ and $\Delta f_\mu$ as
\begin{equation}
\cos\omega\Delta f_e + \sin\omega\Delta f_\mu \leq B(x(\omega),y(\omega),z(\omega))\,,
\end{equation}
with
\begin{align}
x(\omega) &= (1-f_{e,{\rm S}}-2f_{\mu,{\rm S}})\sin\omega\,,\\
y(\omega) &= (1-2f_{e,{\rm S}}-f_{\mu,{\rm S}})\cos\omega\,,\\
z(\omega) &= (f_{\mu, {\rm S}}-f_{e, {\rm S}})(\cos\omega-\sin\omega)\,.
\end{align}
Finally, if we parametrize the electron and muon neutrino flavor shifts in terms of a new parameter $\chi$, as $\Delta f_e = \ell(\chi)\cos\chi$ and $\Delta f_\mu = \ell(\chi)\sin\chi$, we can express the unitarity boundary in flavor space via a boundary on $\ell(\chi)$ as
\begin{equation}\label{eq:boundary}
\ell(\chi) = \min_{\omega}\left\lbrace\frac{ B(x(\omega),y(\omega),z(\omega))}{\cos(\chi-\omega)} \bigg| |\chi-\omega|<\frac{\pi}{2}\right\rbrace\,.
\end{equation} 

In Figure~\ref{fig1} we show the resulting boundaries of Eq.~(\ref{eq:boundary}) of the accessible flavor ratios for three physically motivated choices of flavor composition at the sources --- pion decay $($1\,:\,2\,:\,0)$_{\rm S}$, neutron decay $($1\,:\,0\,:\,0$)_{\rm S}$, and muon-damped pion decay $($0\,:\,1\,:\,0$)_{\rm S}$. Note that the democratic composition $f_{\alpha,\oplus}=1/3$ is always part of the accessible flavor space, independently of the source composition. This occurs when the transition matrix is trimaximal~\footnote{In the PMNS parametrization of the unitary matrix, this can be realized by the mixing angles $\sin\theta_{12}=\sqrt{1/2}$, $\sin\theta_{23}=\sqrt{1/2}$, $\sin\theta_{13}=\sqrt{1/3}$, and Dirac phase $\delta=\pi/2$.}, {\it i.e.,}  ${P}_{\alpha\beta}=1/3$.

By construction, the boundary in Eq.~(\ref{eq:boundary}) encloses a convex subset, {\it i.e.}, one in which every line segment between any two points is contained in the subset. It is a nontrivial question if every flavor combination within the boundary can be actually realized by at least one unitary mixing matrix. In that case, the boundary in Eq.~(\ref{eq:boundary}) would correspond to the convex hull and completely characterize the accessible flavor space. While a general mathematical proof is beyond the scope of this study, we can validate these assumptions for the three benchmark astrophysical compositions shown in Fig.~\ref{fig1} via a numerical analysis. The left and center plots in Fig.~\ref{fig3} show the distribution of observed flavor ratios from 4,000 random realizations of unitary mixing parameters for source compositions $($1\,:\,2\,:\,0)$_{\rm S}$ and $($0\,:\,1\,:\,0$)_{\rm S}$. In both cases, visual inspection of the random samples indicate that the accessible flavor space is convex. This is numerical evidence that Eq.~(\ref{eq:boundary}) completely characterizes the accessible flavor space for the three source compositions shown in Fig.~\ref{fig1}. For comparison, we also show the results previously derived in Ref.~\cite{Xu:2014via} based on a finite set of unitarity bounds.

On the other hand, it is also possible to provide evidence that convexity is not the general case. For instance, the right plot in Fig.~\ref{fig3} shows the distribution of observed flavor ratios from 4,000 random realizations of unitary mixing matrices for $($1\,:\,4\,:\,0)$_{\rm S}$. This source composition could correspond to a neutrino source that has a partially muon-damped composition and, therefore, an enhanced muon neutrino fraction. As before, we also show our convex boundary as a solid line. In this particular case, the distribution is not convex and, therefore, the boundary is not maximally constraining.

The gray-shaded areas in Fig.~\ref{fig1} indicate the 68\% and 95\% confidence levels (C.L.s) from a flavor-composition analysis carried out by IceCube~\cite{Aartsen:2015knd}. Due to the difficulty in distinguishing between events induced by $\nu_e$ and $\nu_\tau$ in the IceCube data~\cite{Abbasi:2012cu,Aartsen:2015dlt}, the likelihood contour is presently rather flat along the $f_\mu$ direction, leading to almost horizontal confidence levels in the ternary plot~\cite{Mena:2014sja, Aartsen:2015ivb, Palomares-Ruiz:2015mka, Vincent:2016nut}. This degeneracy could be lifted in future data by the observation of characteristic $\bar{\nu}_e$~\cite{Glashow:1960zz, Anchordoqui:2004eb, Bhattacharya:2011qu, Barger:2014iua, Palladino:2015vna} and $\nu_\tau$ events~\cite{Learned:1994wg, Aartsen:2014njl, Palladino:2018qgi}. Under the assumption of standard oscillations, the observed flavor composition disfavors the source composition $($1\,:\,0\,:\,0$)_{\rm S}$. However, the unitarity bound indicates that there exist nonstandard oscillation scenarios that can be consistent within the $68\%$ C.L.

In general, we expect that a realistic astrophysical source will be dominated by a source composition that has a low contribution of tau neutrinos, $f_{\tau,{\rm S}} \simeq 0$. The combined unitarity bound of all source compositions $(f_e,1{\rm-}f_e,0)_{\rm S}$ is indicated as the grey-shaded area in the right plot of Fig.~\ref{fig3}. It is simply given by the union of the unitarity boundaries for $($1\,:\,0\,:\,0$)_{\rm S}$ and $($0\,:\,1\,:\,0$)_{\rm S}$. The present 68\% C.L.~shown in Fig.~\ref{fig1} extends beyond this combined region. Therefore, the results of future IceCube analyses with a higher flavor precision have the potential to identify both deviations from standard oscillation and source compositions.

\section{Conclusions}\label{sec3}

The flux of astrophysical neutrinos observed with IceCube allows to test models of neutrino oscillation and interaction at previously inaccessible neutrino energies. In this paper we have discussed general unitarity bounds on the oscillation-averaged flavor composition of high-energy neutrinos emitted by astrophysical sources. These bounds apply to any nonstandard three-flavor neutrino oscillation model where oscillation-averaged flavor transitions are determined by the unitary mixing of flavor and propagation eigenstates. 

We have validated via numerical simulations that our bounds are maximal for typical benchmark source compositions considered in astrophysics and that they allow for a complete characterization of the accessible flavor space. Our focus in this paper was on CP-even effective Hamiltonians that predict the same oscillation phenomena for neutrinos and antineutrinos. The same method can also applied to CP-odd Hamiltonians if one considers the oscillation-averaged flavor compositions of neutrino and antineutrinos separately.

The unitarity bounds allow to study the presence of nonunitary flavor compositions in the astrophysical neutrino data. These compositions could be induced by quantum decoherence~\cite{Hooper:2004xr,Anchordoqui:2005gj,Mehta:2011qb}, sterile neutrinos~\cite{Beacom:2003eu, Keranen:2003xd, Esmaili:2009fk,Hollander:2013im,Shoemaker:2015qul,Brdar:2016thq}, neutrino decay~\cite{Beacom:2002vi,Barenboim:2003jm,Maltoni:2008jr,Mehta:2011qb,Baerwald:2012kc,Pagliaroli:2015rca,Shoemaker:2015qul,Bustamante:2016ciw,Denton:2018aml}, extra dimensions~\cite{Lykken:2007kp,Kisselev:2010zz,Aeikens:2014yga,Esmaili:2014esa} or inelastic scattering in the cosmic neutrino background~\cite{Ioka:2014kca, Ng:2014pca, DiFranzo:2015qea} or dark matter~\cite{Reynoso:2016hjr,Arguelles:2017atb}. We refer to the recent study~\cite{Rasmussen:2017ert} for a detailed discussion.

Production of tau neutrinos in astrophysical sources is expected to be strongly suppressed. Under this assumption, we have derived a region in the observable neutrino flavor space that cannot be accessed by astrophysical sources if oscillations respect unitarity. Presently, the flavor composition based on IceCube data is consistent with the standard oscillation predictions. However, the 68\% confidence region allows for other flavor compositions generated by nonstandard oscillations or source compositions.

We have provided a refined and streamlined formalism to derive unitarity bounds that are easily applicable to arbitrary source compositions. In doing so, we have elevated unitarity bounds to being useful tools for future searches of new physics in astrophysical neutrinos.

\acknowledgements
We would like to thank Guoyuan Huang for useful comments on our manuscript. M.A.~and M.B.~acknowledge support from Danmarks Grundforskningsfond (project no.~1041811001) and \textsc{Villum Fonden} (projects no.~18994 and no.~13164, respectively). S.M.~would like to thank Thomas Greve and the Caltech SURF program for financial support.

\appendix

\section{Unitarity Bounds}\label{appI}

For the derivation of the boundary function of Eq.~(\ref{eq:bound}) we follow the procedure outlined in Ref.~\cite{Xu:2014via}. The oscillation-averaged neutrino flavor-transition matrix can be written as the matrix product
\begin{equation}\label{eq:PQQ}
{\bf P}={\bf Q}{\bf Q}^T\,,
\end{equation}
where ${Q}_{\alpha i} \equiv |U_{\alpha i}|^2$. The matrix elements of ${\bf Q}$ are subject to the unitarity condition ${\bf U}^\dagger{\bf U} = {\boldsymbol 1}$. This imposes the normalization condition $\sum_\alpha{Q}_{\alpha i}=1$ and the boundary condition
\begin{equation}\label{eq:bound1}
0\leq{Q}_{\alpha i}\leq1\,.
\end{equation}
In addition, the elements of ${\bf Q}$ are subject to triangle inequalities that can be summarized by the condition
\begin{equation}\label{eq:bound2}
T(\sqrt{Q_{\alpha 1}Q_{\beta 1}},\sqrt{Q_{\alpha 2}Q_{\beta 2}},\sqrt{Q_{\alpha 3}Q_{\beta 3}})\geq0\,,
\end{equation}
where the function $T$ is defined as
\begin{multline}\label{eq:T}
T(a,b,c) \equiv (a+b+c)(a+b-c)\\\times(b+c-a)(c+a-b)\,,
\end{multline}
and is proportional to the squared area of a triangle with sides $a$, $b$, and $c$. 

\begin{table}[t]\centering
\begin{minipage}[t]{\linewidth}
\begin{ruledtabular}
\begin{tabular}{ccc}
Value & ${\bf Q}$ & Condition\\[0.1cm]
\hline
\begin{minipage}[c][1.5cm][c]{0.33\columnwidth}$0$\end{minipage} & 
$\begin{pmatrix}
1&0&0\\
0&1&0\\
0&0&1
\end{pmatrix}$&
---\\
\begin{minipage}[c][1.5cm][c]{0.33\columnwidth}$\displaystyle\frac{x+y+z}{3}$\end{minipage}  & 
$\begin{pmatrix}
\sfrac{1}{3}&\sfrac{1}{3}&\sfrac{1}{3}\\
\sfrac{1}{3}&\sfrac{1}{3}&\sfrac{1}{3}\\
\sfrac{1}{3}&\sfrac{1}{3}&\sfrac{1}{3}
\end{pmatrix}$ & 
---\\
\begin{minipage}[c][1.5cm][c]{0.33\columnwidth}$\displaystyle\frac{x}{2}$\end{minipage} &
$\begin{pmatrix}
1&0&0\\
0&\sfrac{1}{2}&\sfrac{1}{2}\\
0&\sfrac{1}{2}&\sfrac{1}{2}
\end{pmatrix}$ &   
---\\
\begin{minipage}[c][1.5cm][c]{0.33\columnwidth}$\displaystyle\frac{(3x+y+z)^2-4yz}{24x}$\end{minipage} &
$\begin{pmatrix}
0&p&1{\rm -}p\\
\sfrac{1}{2}&\frac{1-p}{2}&\frac{p}{2}\\
\sfrac{1}{2}&\frac{1-p}{2}&\frac{p}{2}
\end{pmatrix}$ &  
\begin{minipage}{3cm}$\displaystyle0\leq p\leq1$ \\[0.3cm]
$\displaystyle p\equiv\frac{3x+y-z}{6x}$\end{minipage}\\
\end{tabular}
\end{ruledtabular}
\end{minipage}
\caption[]{Classes of candidate maxima of the function $G({\bf Q};x,y,z)$. The full set of candidates can be recovered by applying the transformations in Eq.~(\ref{eq:invariance}).}\label{tab1}
\end{table}

The bound $B(x,y,z)$ in Eq.~(\ref{eq:bound}) corresponds to the global maximum of the function
\begin{equation}\label{eq:G}
G({\bf Q};x,y,z) = x{P}_{\mu\tau} + y {P}_{e\tau} + z{P}_{e\mu}
\end{equation}
for all possible choices of ${\bf Q}$. We follow the procedure outlined in Ref.~\cite{Xu:2014via} by first identifying all possible extrema $\mathcal{S}(x,y,z)$ of Eq.~(\ref{eq:G}) and selecting the global maximum as in Eq.~(\ref{eq:B}). Due to the normalization condition $\sum_\alpha{Q}_{\alpha i}=1$, we can maximize $G$ with respect to, say, $(Q_{e 1},Q_{e 2},Q_{\mu 1},Q_{\mu 2})$, subject to the boundary conditions, Eqs.~(\ref{eq:bound1}) and (\ref{eq:bound2}). Before we proceed, we note two simplifications of our approach compared to the method outlined in Ref.~\cite{Xu:2014via}: 
\begin{enumerate}[\rm (i)]
\item The set of local extrema $\mathcal{S}(x,y,z)$ of Eq.~(\ref{eq:G}) is invariant under the transformation
\begin{equation}\label{eq:invariance}
Q'_{\alpha i } = Q'_{s_\alpha \bar{s}_i}\qquad x'_i = x_{\bar{s}_i}\,,
\end{equation}
where $s$ and $\bar{s}$ are two permutations of the indices and ${\bf x} \equiv (x,y,z)$. In other words, the solutions are invariant under exchange of entries of two arbitrary columns of the matrix ${\bf Q}$ or the simultaneous exchange of rows and parameters $x$, $y$, and $z$. In the following, we will therefore only derive solutions that are not related by the transformation (\ref{eq:invariance}). The final list of candidate extrema in Eq.~(\ref{eq:B}) can then be recovered by applying these transformations.
\item The boundary conditions set by the triangle inequalities in Eq.~(\ref{eq:bound2}) can only be satisfied if there is at least one matrix element of ${\bf Q}$ equal to zero. One can show this by studying the extrema of the function in Eq.~(\ref{eq:T}). All solutions require at least one entry with $Q_{\alpha i}=0$, except for one single solution where $Q_{\alpha 1}=1/3$ for all entries. However, this last extremum is a maximum. With this observation, we only need to identify local extrema of $G$ along the boundary condition $Q_{\alpha i}=0$ for at least one matrix element and do not need to include the surface $T=0$ via a Lagrange multiplier as done in Ref.~\cite{Xu:2014via}.
\end{enumerate}
Table \ref{tab1} lists the four classes of candidate extrema up to transformations described by Eq.~(\ref{eq:invariance}). Except for the second extremum, $Q_{\alpha i}= 1/3$, all candidates can be found on the boundary with at least one entry $Q_{\alpha i}= 0$.

\bibliographystyle{utphys_mod}
\bibliography{references}

\end{document}